\documentclass[journal]{IEEEtran}
\usepackage[pdftex]{graphicx}
\usepackage{amsmath}
\ifCLASSINFOpdf
\else
\fi
\begin{document}
%
\title{Multifunctional Metasurface: Simultaneous Beam Steering, Polarization Conversion and Phase Offset}
%
%
%

\author{Xiaozhen Yang, Erda Wen, Dinesh Bharadia,~\IEEEmembership{Member,~IEEE} and Daniel F. Sievenpiper,~\IEEEmembership{Fellow,~IEEE}
\thanks{This work is supported by National Science Foundation under Grant No.2107613.}
\thanks{The authors are with the Electrical and Computer Engineering Department,
University of California, San Diego, La Jolla, CA 92093-0021 USA (e-mail: xiy003@eng.ucsd.edu; ewen@eng.ucsd.edu; dineshb@eng.ucsd.edu; dsievenpiper@eng.ucsd.edu).}}

\markboth{Multifunctional metasurface}%
{Shell \MakeLowercase{\textit{et al.}}: Bare Demo of IEEEtran.cls for IEEE Journals}
%


\maketitle

\begin{abstract}
A varactor-based reconfigurable multifunctional metasurface capable of simultaneous beam steering, polarization conversion and phase offset is proposed in this paper. The unit cell is designed to naturally decompose the incident waves into two equal amplitude orthogonal linear components, and by integrating varactors, the reflection phase of the field components can be engineered from $-180^{\circ}$ to $180^{\circ}$.Taking advantage of the infinite states of the varactors, this design integrates a new function, the phase offset. After simulation validation of its capability, a four-layer $7$ by $6$ unit one-dimensional prototype is fabricated as a printed circuit board. It is experimentally demonstrated that it switches between X/Y and circular polarization with more than $10$ dB cross polarization isolation, while reaching $\pm45^{\circ}$ steering and $\pm180^{\circ}$ phase offset.
\end{abstract}


%
\IEEEpeerreviewmaketitle

\section{Introduction} \label{Introduction}

\IEEEPARstart{A}{metasurface} \cite{Li18_R,Hu21_R}, the two-dimensional (2-D) counterpart of a metamaterial, exhibits unprecedented abilities of wave manipulation such as polarization conversion, beam steering and phase control. Besides, compared to a 3-D metamterial, its low-profile, low-cost and easy fabrication features make it more appealing. It has aroused wide attention from the microwave regime, to the terahertz and optical ranges for its promising applications in wireless systems to improve robustness and efficiency of communication between the transmitter and receiver, especially for modern ultra-compact electronic systems which are usually linearly-polarized \cite{Ahmed23_R,Cui19_R}. The polarization mismatch can be alleviate by polarization conversion to match with the polarization of the direct path. By beam steering, the maximum amount of power is redirected to the receiver \cite{Dunna20_MIMO}. The reflected beam of a reflective metasurface can constructively combine with direct path by phase matching. In addition to that, such a surface can be a useful attachment to the stop sign, for retro-reflecting the signal back to radar, the polarization change would enable detection of the stop sign in the crowded reflection from the rest of the environment \cite{Bansal22_SS}. We believe developing a multifunctional metasurface with simultaneous polarization, beam steering and phase offset can find many applications in wireless communication systems.

\begin{figure}[!t]
\centering
\includegraphics[width=3.5in]{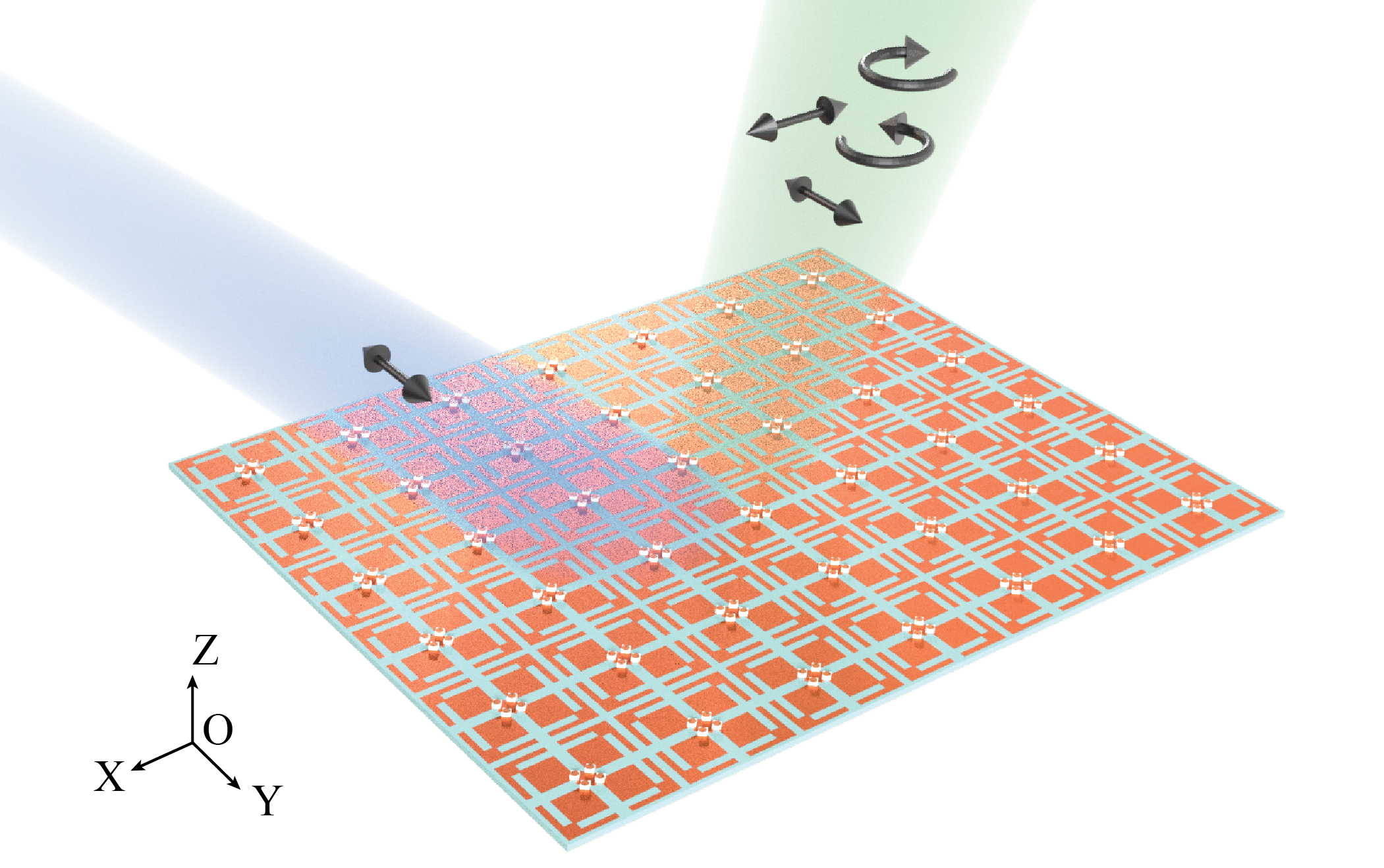}
\caption{Illustration of the proposed simultaneous multifunctional metasurface. Blue beam: incident wave. Green beam: outgoing beam.}
\label{concept} 
\end{figure}

However, conventional passive metasurfaces is only capable of one specific polarization conversion function, for example, \cite{Zhu13_PC,Sun17_PC,Khan19_PC,He22_PC,Huang22_PC} propose to adopt an anisotropic design which retards one of the orthogonal electric field components to achieve linear (LP) to cross-linear (C-LP) or circular polarization (CP) conversion. It is also noticed that the polarization conversion is not optional, which hinders their applications in the real world. Recent research \cite{Deng21_Plasma,Ding18_Plasma} report passive designs realizing simultaneous beam steering and polarization conversion on a gap-surface plasmon metasurface \cite{McDonnell22_THz,Ding18_R,Ding22_R}. Since its electromagnetic response is dependent on the geometry and alignment of the meta-atom, designs have to be modified when aiming at a different polarization conversion or steering angle, limiting its realistic applications.

Studies on reconfigurable metasurfaces \cite{Zahra21_R,He19_R,Fu20_BS,Hwang22_TA,Ni17_PC} address this issue by integrating nonlinear components or materials to increase the versatility. \cite{Sievenpiper03_BS} demonstrates a beam steering capable 2-D metasurface, though polarization conversion is absent, by deploying biased varactors based on high impedance surface \cite{Sievenpiper99_HIS}. By varying the biasing voltage across the varactors, a resonant frequency shift occurs and the reflection phase of each unit cell changes accordingly, achieving a 2-D $\pm 40^{\circ}$ steering. Some works \cite{Magarotto23_Plasma,Wang22_THz} adopt magnetized plasma and biased ferromagnetic material, respectively, to achieve simultaneous beam steering and polarization control. However, the employment of ferromagnetic material and external biasing field using magnets or coils can lead to a heavy and bulky design. Recent research on transmitarrays and reflectarrays use diodes to switch between different polarizations and phase delay \cite{Rana21_TA,Huang15_TA}, limited by their 1-bit phase resolution, these works are only capable of simultaneous beam steering and LP to LP/C-LP conversion. A reflectarray study \cite{Hu22_RA} realizes LP to LP/CP, using a dipole and patch unit cell. These p-i-n diode-based approach requires a very large surface, usually above $10$ by $10$-units, since one diode can only switch between two reflection phases, and using a small surface results in inaccurate steering angle. Moreover, those unit cell designs are electrically large (above $0.4\lambda$). 

Besides the metasurface approach discussed above, an antenna array is also a potential candidate, but it suffers from its large size \cite{Hu21_AA} compared with metasurfaces, whose unit cell periodicity is usually below a half wavelength($1/2\:\lambda$), and the thickness of the whole design is around $1/50\:\lambda$.

Except for polarization conversion and beam steering, another function, adding a phase offset, is potentially useful but is not reported in other reconfigurable metasurface works. This function enables phase manipulation of the reflected beam, such that a spatial hot/cold spot could be created when using more than one metasurface to cancel or enhance the local field.

In this article, we propose the first multifunctional reflective metasurface, as illustrated in Fig.\ref{concept}, with simultaneous beam steering, polarization conversion including X/Y polarization to X/Y/LHCP/RHCP conversion, and complete phase offset range from $-180^{\circ}$ to $180^{\circ}$ by adopting biased varactors. A unit cell which responds to the two orthogonal linear-polarized components of the incident wave is designed. Deploying varactors allows phase manipulation on the orthogonal components to achieve polarization conversion within unit cells. An overall phase shift across adjacent unit cells enables beam steering. The phase offset is realized by assigning different initial phase values to the first column of unit cells. 

\begin{figure}[!t]
\centering
\includegraphics[width=3.5in]{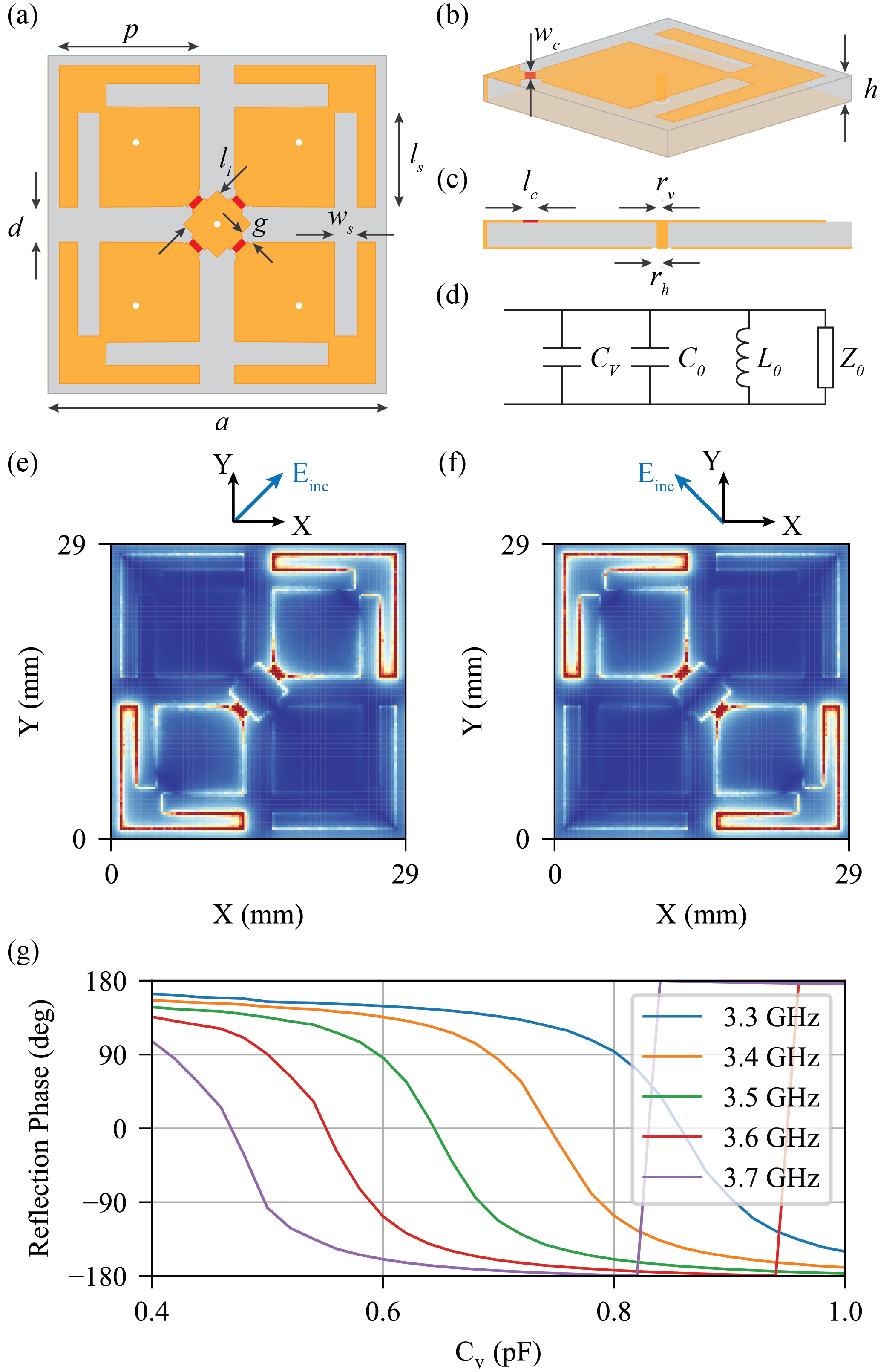}
\caption{Unit cell illustration and simulated results. (a) Top view of the unit cell. (b) and (c) Side veiw of a quarter unit cell. (d) Equivalent circuit of a quarter-cell with a varactor. Substrate: Duroid Rogers 5880. (e) and (d): field distribution under two orthogonal incident, respectively. (g) Reflection phase of the unit cell with various $C_v$.}
\label{HFSS-Unit} 
\end{figure}

The rest of the paper is arranged in the following way. In section \ref{HFSS}, the unit cell design and its working mechanism is discussed, and as a proof of concept, simulation results of a $7$ by $6$-unit surface are also presented. Experimental results are shown in Section \ref{Meas.}. Finally, the conclusions are reached in Section \ref{Conclusion}. 

\section{Unit Cell Design and Simulation Results} \label{HFSS}
\subsection{Unit Cell Design and Mechanism}

Any LP can be decomposed into two orthogonal LPs, take $y$-polarized wave as an example,
\begin{equation}
    \hat{y}Ee^{j\phi_{0}} = \frac{\hat{x}+\hat{y}}{\sqrt{2}}\frac{E}{\sqrt{2}}e^{j\phi_{1}}+\frac{\hat{y}-\hat{x}}{\sqrt{2}}\frac{E}{\sqrt{2}}e^{j\phi_{2}},
\label{eqn1}
\end{equation}
where $\phi_{1,2}=\phi_0$. If the phase of the orthogonal components can be manipulated properly, polarization conversion is achieved. LP to LP/CP/C-LP conversion is realized by simply adding on an extra phase delay to the components in Eqn.(\ref{eqn1}) when $\phi_1 - \phi_2 = 0^{\circ}/\pm90^{\circ}/\pm180^{\circ}$, otherwise, LP to EP conversion.

To realize this, we first design a unit cell structure which naturally decomposes the X/Y polarization to two orthogonal components with independent electromagnetic response. As illustrated in Fig.\ref{HFSS-Unit} (a) to (c), the unit consists of four identical arrow-shape patches with ungrounded vias underneath and a grounded square island at the center. The slots on the patch are designed to shrink the electrical size for the resonance. The dimensions of the structure are provided in Table.\ref{Table-unit}. The locations of the ungrounded vias are chosen to be the weakest field distribution points, so that the vias do not interfere with the resonance and this can reduce the leakage to the biasing system. When illuminated by a Y-polarized electric field, the patches on the two diagonals react independently to the orthogonal components of the incident wave, as shown in Fig.\ref{HFSS-Unit} (e) and (f), demonstrating its capability to respectively modify the reflection phase of the field components to accomplish polarization conversion. 

\begin{table}[!ht]
    \centering
    \caption{Dimensions of the unit cell design (Unit: mm)}
    \begin{tabular}{c|cccccc}
    \hline
        Parameter & $a$ & $p$ & $d$ & $g$ & $h$ & $l_i$ \\ \hline
        Value & 29 & 12 & 3 & 1.2 & 1.575 & 4 \\ \hline
        Parameter & $w_s$ & $l_s$ & $w_c$ & $l_c$ & $r_v$ & $r_h$ \\ \hline
        Value & 2 & 8 & 1.2 & 0.58 & 0.25& 0.50 \\ \hline
    \end{tabular}
    \label{Table-unit}
\end{table}

Reflection phase manipulation of this design relies on varactors, illustrated as the red rectangles in Fig.\ref{HFSS-Unit} (a). Four varactors (MACOM MA46H070-1056, $0.3$ pF to $1.1$ pF) are applied in parallel to the metal patches across the gap between patches and the central island, creating an equivalent circuit shown in (d) for one quarter of the unit. The resonating frequency is determined by
\begin{equation}
    f_0 = \frac{1}{2\pi}\sqrt{\frac{1}{L_0 (C_0+C_v)}},
\end{equation}
where $L_0$ and $C_0$ is the intrinsic inductance and capacitance of the patch, and $C_v$ is the capacitance introduced by a varactor. When applying different voltage through the ungrounded vias, a frequency shift on $f_0$ occurs, resulting in a reflection phase change. To study the reflection phase, all varactors are assigned with the same $C_v$ in the unit cell simulation with periodic boundaries. All simulations are performed in Ansys HFSS and the results are shown in Fig.\ref{HFSS-Unit} (g) with various $C_v$. Due to the lack of a SPICE model and for simplicity, the varactors are treated as lossless capacitors in the simulations and the biasing network is neglected. The relationship between capacitance and biasing voltage is provided in the product datasheet by the manufacturer. A complete reflection phase range from $-180^{\circ}$ to $180^{\circ}$ is observed from $3.3$ GHz to $3.7$ GHz within the varactors capacitance range under normal incidence for X/Y polarization, proving its capability of reflection phase manipulation. To activate polarization conversion, varactors on the same diagonals should be assigned with the same capacitance, while the difference between the two diagonals should be chosen properly for different conversions.

\begin{figure}[!t] 
\centering
\includegraphics[width=3.5in]{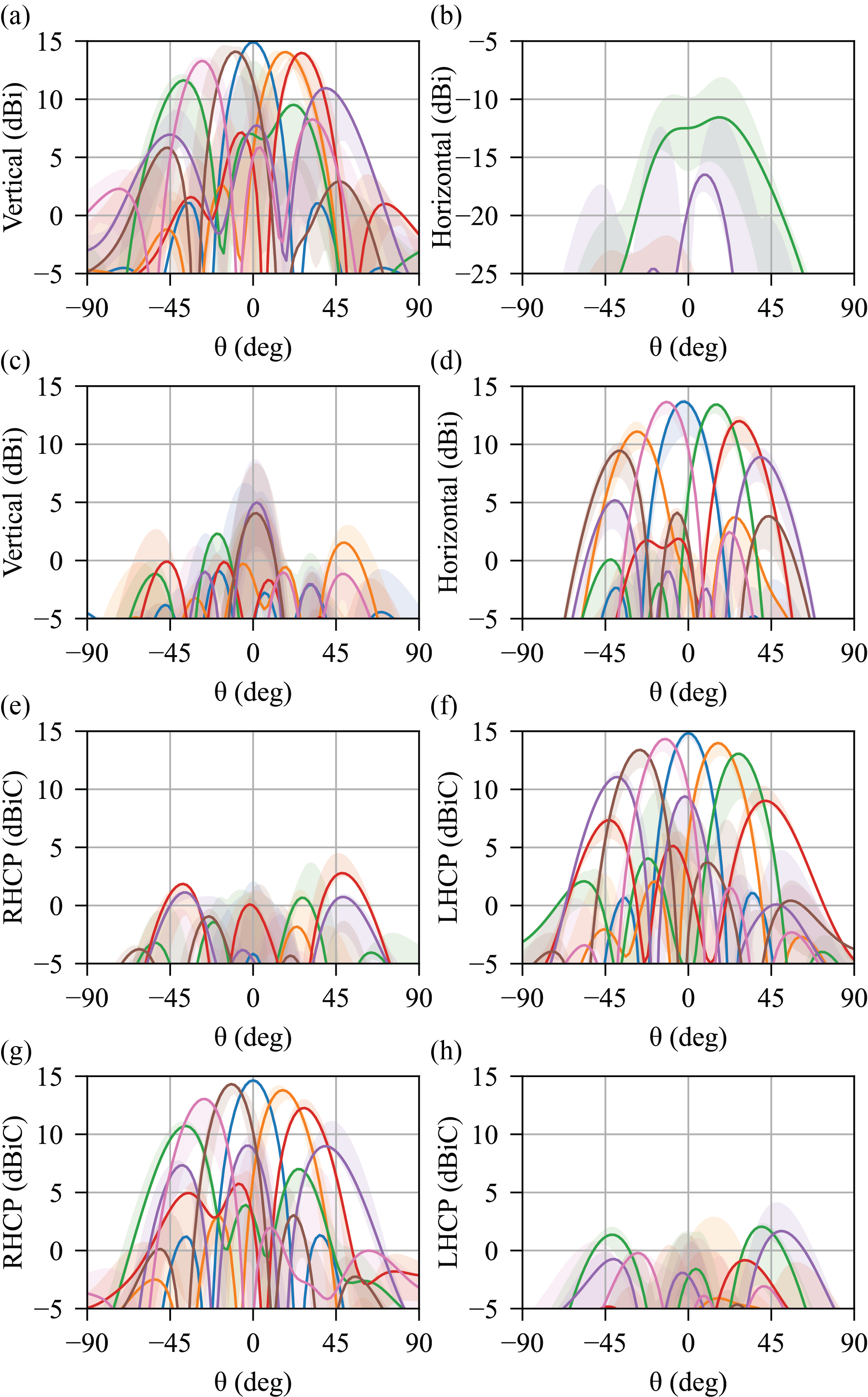}
\caption{Cross-polarization isolation analysis for the $7$ by $6$-unit metasurface design under normal $y$-polarized incidence. The design frequency is $3.6$ GHz with $\phi_0=150^{\circ}$. Solid line: $3.6$ GHz, shaded area:$3.58$ GHz to $3.63$ GHz. (a) and (b): LP ($x$) to LP ($x$) conversion. (c) and (d): LP to C-LP ($y$) conversion. (e) and (f): LP to RHCP conversion. (g) and (h): LP to LHCP conversion.}
\label{HFSS-Pattern} 
\end{figure}

\begin{figure}[!t] 
\centering
\includegraphics[width=3.5in]{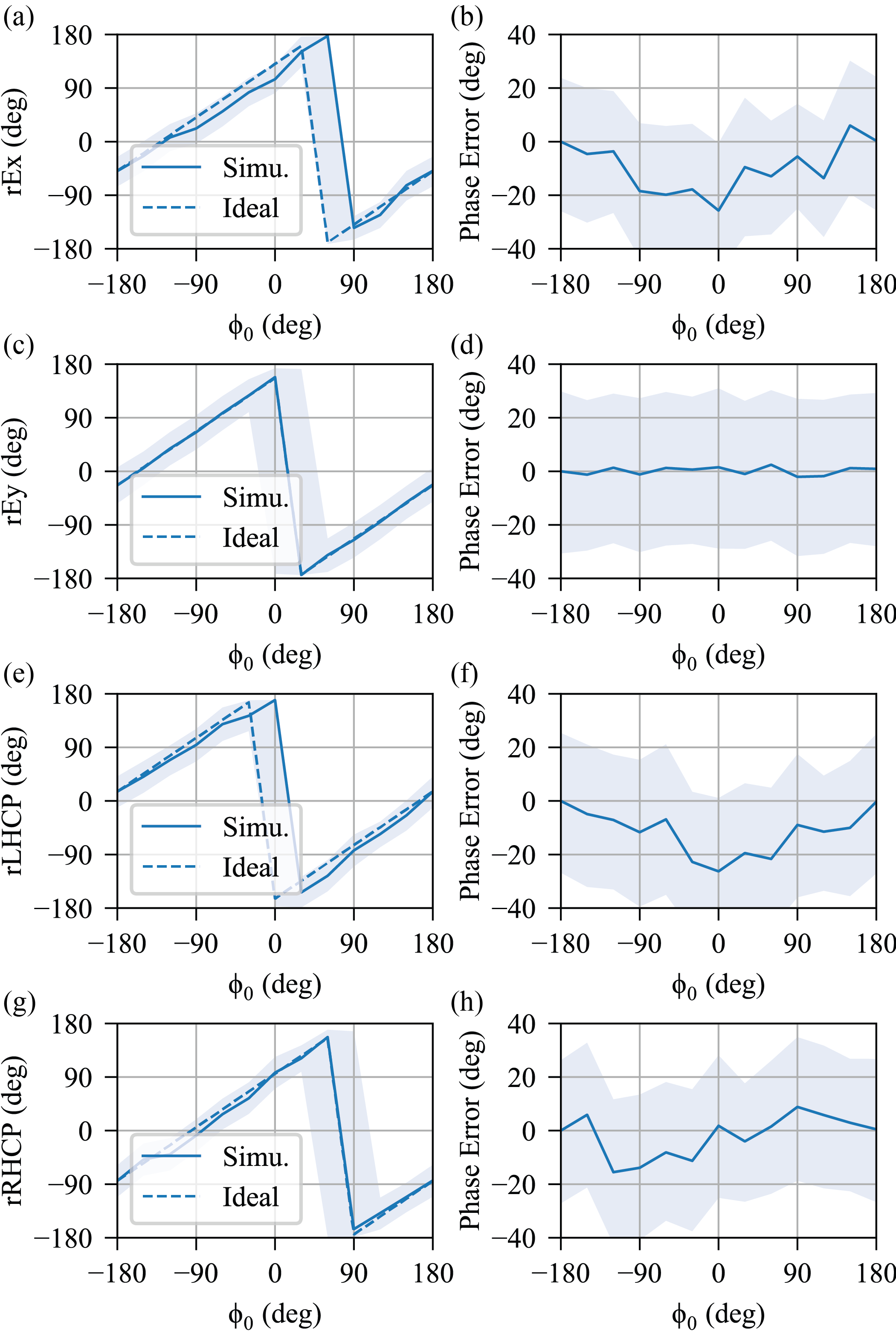}
\caption{The phase offset added to the reflected beam with various $\phi_0$, for $30^{\circ}$ steering at $3.6$ GHz. Solid line: main beam direction, shaded area: $\pm 5^{\circ}$ of the main beam. (a) and (b): LP ($x$) to LP ($x$) conversion. (c) and (d): LP to C-LP ($y$) conversion. (e) and (f): LP to LHCP conversion. (g) and (h): LP to RHCP conversion.}
\label{HFSS-Offset} 
\end{figure}

Beam steering is realized by a phase difference across adjacent units. An external phase difference should be added on to the next neighboring unit while maintaining the internal phase difference between the two diagonals for simultaneous beam steering and polarization conversion. Assume the target is to convert a normal X-polarized incident to a Y-polarized outgoing wave while steering the main lobe by $\theta = 30^{\circ}$. The external phase delay between units should be
\begin{equation}
    \Delta \phi_e = a\:\sin (\theta)\:k = 62.6^{\circ},
\end{equation}

where $k$ is the wave number and $a$ is the periodicity of the unit cell. If the phase of the first unit is $\phi^1_1=180^{\circ}$ and $\phi^1_2=0^{\circ}$ for the two diagnoals, respectively, the phase for the second unit should be $\phi^2_1=\phi^1_1-\Delta\phi_e=117.4^{\circ}$ and $\phi^2_2=\phi^1_2-\Delta\phi_e-62.6^{\circ}$, the third unit should take $\phi^3_1=\phi^2_1-\Delta\phi_e=54.8^{\circ}$ and $\phi^3_2=\phi^2_2-\Delta\phi_e=-125.2^{\circ}$, where $\phi^n_1$ is the reflection phase of the upper-left to lower-right diagonal of the $n^{th}$ unit and $\phi^n_2$ is the reflection phase of the upper-right to lower-left diagonal of the $n^{th}$ unit.

There are numerous combinations of $\phi^n_1$ and $\phi^n_2$ that result in the same steer angle and polarization conversion, since the phase of the first unit cell (referred to as the initial phase in this paper), $\phi_0=\phi^1_1$ , is free of choice from $-180^{\circ}$ to $180^{\circ}$, which gives the third function of the proposed structure: phase offset. As $\phi_0$'s range is complete, the phase of the outgoing beam is also complete from $-180^{\circ}$ to $180^{\circ}$.

In conclusion, polarization conversion is achieved by reflection phase difference between the two diagonals, $\Delta\phi_i=\phi^n_1-\phi^n_2$, beam steering is realized by external phase difference $\Delta \phi_e$ across units, and the phase offset function is reached by the different choice of the initial phase $\phi_0$.

\subsection{Extending to Metasurface}
The unit cell is then extended to a $7$ by $6$ unit metasurface, as shown in Fig.\ref{concept}, to prove its simultaneous multifunctional capability in driven modal simulations. For simplicity, each column of the surface is assigned with the same capacitance for the two diagonals, respectively, to achieve 1-D steering in the XOZ plane. Four cases are studied under plane wave normal incidence: $Y$-polarized excitation to LP/C-LP/RHCP/LHCP conversion with steering angle varying from $-45^{\circ}$ to $45^{\circ}$. As shown in Fig.\ref{HFSS-Pattern}, the cross-polarization analysis demonstrates approximately $10$ dB isolation between the target polarization and its orthogonal while simultaneously performing beam steering. In this figure, all results are obtained using $\phi_0=150^{\circ}$ and the design frequency is $3.6$ GHz, using the red curve in Fig.\ref{HFSS-Unit} (g). The shaded area in Fig.\ref{HFSS-Pattern} denotes the bandwidth from $3.58$ GHz to $3.63$ GHz. The capacitance used for LP to LHCP conversion is provided in Table.\ref{Table-cap}, where $C^n_{1,2}$ corresponds to the capacitance for $\phi^n_{1,2}$. The programmable and reconfigurable features of this design enable a wide possible operational frequency range from $3.3$ GHz to $3.7$ GHz, since these frequencies also present a complete reflection phase range from $-180^{\circ}$ to $180^{\circ}$, as shown in Fig.\ref{HFSS-Unit} (g). The simulated phase offset by varying $\phi_0$ is illustrated in Fig.\ref{HFSS-Offset}. The simulated phase offset $\phi_{rE}^{Simu.}$ is defined by the phase of the radiated far field,
\begin{equation}
    rE = |rE|e^{\phi_{rE}^{Simu.}}.
\end{equation}
The ideal phase offset if calculated by
\begin{equation}
    \phi_{rE}^{Ideal} = \phi_{rE}^{Simu.}(\phi_0=-180^{\circ})+\phi_0,
\end{equation}
which assumes that any variation in $\phi_0$ results in the same amount of variation in the phase of the radiated far field. Due to the imperfect choice of $C_v^n$, a phase error between $\phi_{rE}^{Simu.}$ and $\phi_{rE}^{Ideal}$ occurs. A $\pm20^{\circ}$ accuracy is observed for the main beam direction, and the phase variation for $\pm5^{\circ}$ around the main beam is $\pm40^{\circ}$.

\begin{table}[!ht]
    \centering
    \caption{Capacitance used on each channel for LP to LHCP conversion.\\(Unit: pF)}
    \begin{tabular}{c|ccccccc}
    \hline
        Steering angle & $C^1_1$ & $C^1_2$ & $C^2_1$ & $C^2_2$ & $C^3_1$ & $C^3_2$ & $C^4_1$ \\ \hline
        $-45^{\circ}$ & 0.30 & 0.52 & 0.52 & 0.56 & 0.56 & 0.61 & 0.61 \\ 
        $-30^{\circ}$ & 0.30 & 0.52 & 0.51 & 0.55 & 0.54 & 0.58 & 0.57 \\ 
        $-15^{\circ}$ & 0.30 & 0.52 & 0.47 & 0.54 & 0.51 & 0.55 & 0.53 \\ 
        $0^{\circ}$ & 0.30 & 0.52 & 0.30 & 0.52 & 0.30 & 0.52 & 0.30 \\ 
        $15^{\circ}$ & 0.30 & 0.52 & 0.94 & 0.50 & 0.65 & 0.45 & 0.61 \\ 
        $30^{\circ}$ & 0.30 & 0.52 & 0.66 & 0.46 & 0.59 & 0.86 & 0.56 \\ 
        $45^{\circ}$ & 0.30 & 0.52 & 0.62 & 0.32 & 0.56 & 0.62 & 0.53 \\ \hline
        Steering angle & $C^4_2$ & $C^5_1$ & $C^5_2$ & $C^6_1$ & $C^6_2$ & $C^7_1$ & $C^7_2$ \\ \hline
        $-45^{\circ}$ & 0.30 & 0.30 & 0.52 & 0.52 & 0.56 & 0.56 & 0.61 \\ 
        $-30^{\circ}$ & 0.62 & 0.60 & 0.73 & 0.73 & 0.49 & 0.41 & 0.53 \\ 
        $-15^{\circ}$ & 0.57 & 0.55 & 0.56 & 0.56 & 0.60 & 0.57 & 0.63 \\ 
        $0^{\circ}$ & 0.52 & 0.30 & 0.52 & 0.30 & 0.52 & 0.30 & 0.52 \\ 
        $15^{\circ}$ & 0.30 & 0.58 & 0.57 & 0.57 & 0.61 & 0.56 & 0.60 \\ 
        $30^{\circ}$ & 0.61 & 0.54 & 0.49 & 0.49 & 0.55 & 1.30 & 0.51 \\ 
        $45^{\circ}$ & 0.56 & 0.36 & 0.62 & 0.62 & 0.37 & 0.57 & 0.63 \\ \hline
    \end{tabular}
    \label{Table-cap}
\end{table}

\section{Measurement and Results} \label{Meas.}

\subsection{Fabrication and Measurement Setup}
\begin{figure}[!t]
\centering
\includegraphics[width=3.5in]{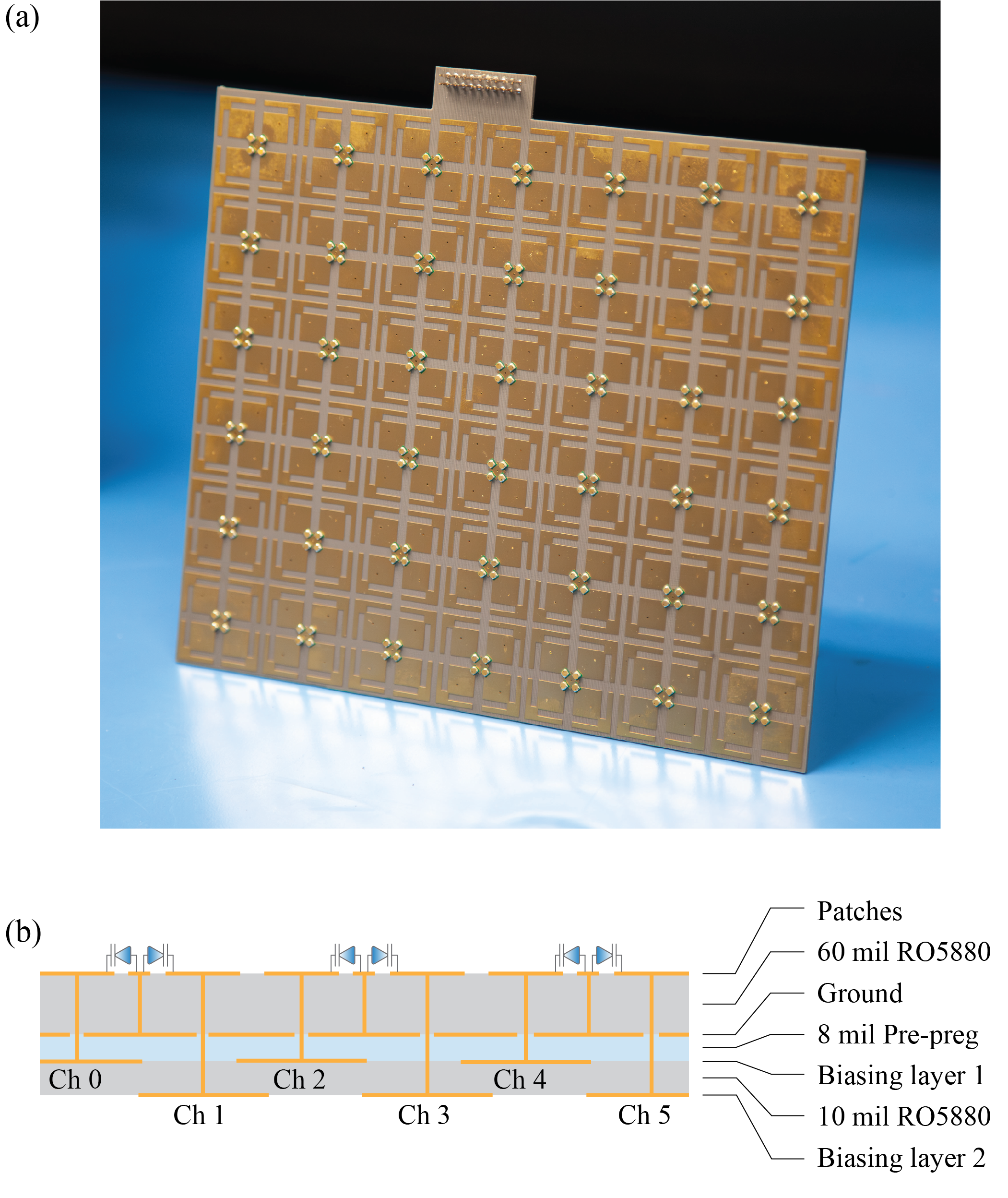}
\caption{The fabricated protoype. (a) A picture of the $7$ by $6$-unit metasurface. (b) The stack-up illustration including the biasing layers.}
\label{Fab} 
\end{figure}

\begin{figure}[!t]
\centering
\includegraphics[width=3.5in]{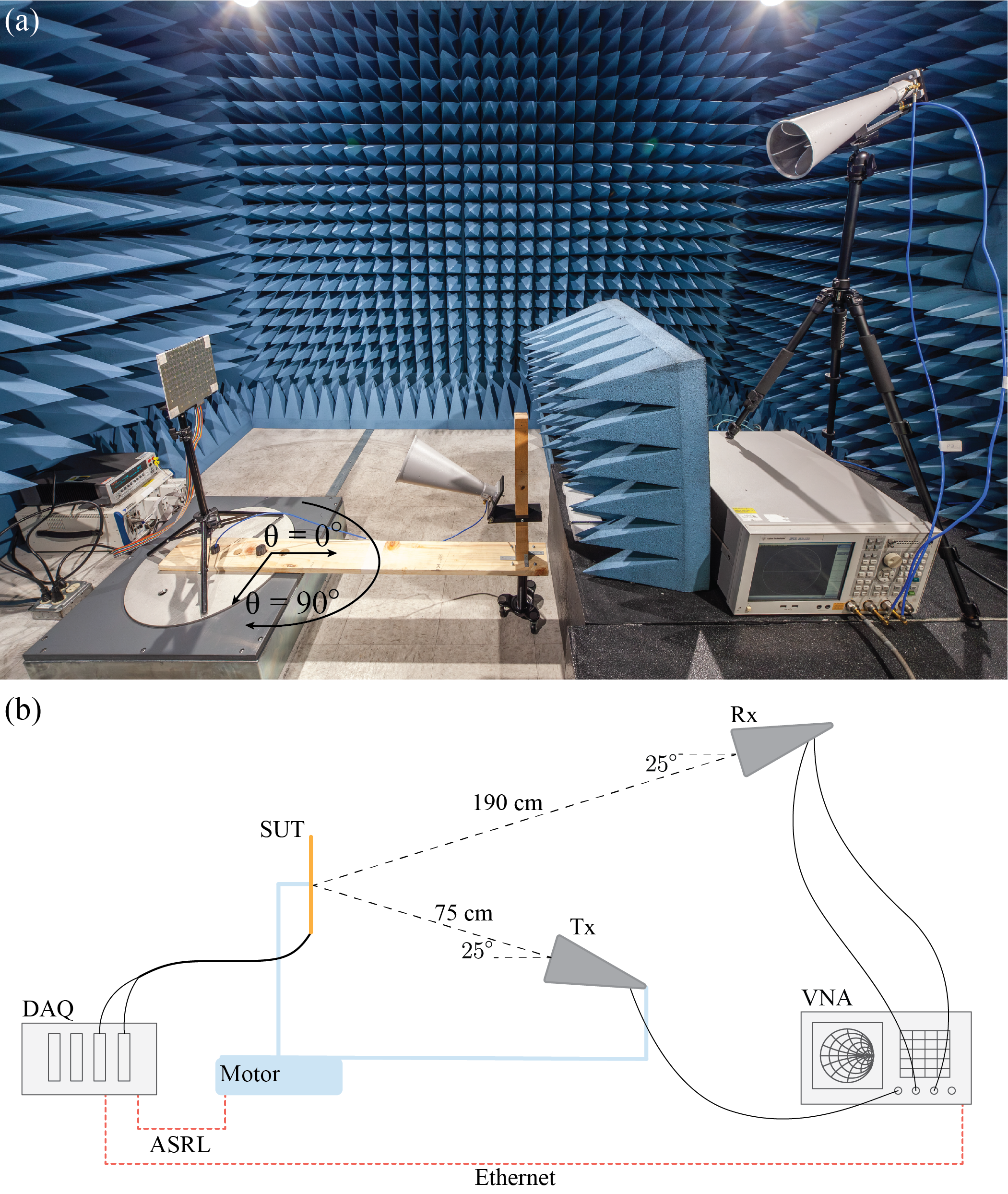}
\caption{Experiment setup for reflection phase and pattern measurement. (a) A picture of the measurement setup in the anechoic chamber. (b) An illustration of all equipment and their connection.}
\label{Meas.-Setup} 
\end{figure}

After numerical validations for the proposed structure, a varactor-loaded, four-layer, $7$ by $6$-unit 1-D surface is manufactured as shown in Fig.\ref{Fab} (a). This prototype consists of two Duroid Rogers 5880 cores, bonded by a $8$ mil insulating pre-preg material. The total thickness of this board is around $80$ mil. The top two layers are the varactor-integrated patches and the ground plane, respectively. Underneath is a $10$ mil thick double-layer DC biasing network. Each unit has two biasing channels, corresponding to the patches on the two diagonals, and the units on each column are connected together, resulting in 14 DC channels, as illustrated in Fig.\ref{Fab} (b). To avoid intersection of the biasing lines, all the wires for the upper-left to lower-right diagonals are fabricated on biasing layer 1, and the upper-right to lower-left ones are on biasing layer 2.

The experiment setup for all measurements is shown in Fig.\ref{Meas.-Setup}. The surface under test (SUT) is fixed on a tripod, sitting on a motor which rotates in azimuth. The surface is biased by a data acquisition device (DAQ, NI PXIe-1062Q) through ribbon cables. The DAQ is used as the host computer for all devices to realize unsupervised data collection controlled by Python. The transmitter (Tx) and receiver (Rx) are identical dual-polarization horn antennas (RCDLPHA2G18B) and thus can be used to measure circularly-polarized waves using both channels on Rx. The Tx, connected to the vector network analyzer (VNA, E5071C), is tilted by $25^{\circ}$ upwards and fixed on a wooden frame, which rotates simultaneously with the motor. Using the VNA as the source, the incident wave is radiated by Tx, reflected by the SUT and then collected by both channels of Rx, which are also connected to the VNA. Three channels of the VNA are involved in the experiment, the first port excites Tx's vertical channel, the second and third port are connected to the vertical and horizontal channel of Rx, respectively. Thus, $S_{21}\:(S_{vv})$ and $S_{31}\:(S_{hv})$ describes the the LP to LP and LP to C-LP conversion rate. The distance between the Tx and SUT is around $75$ cm for plane wave illumination. The Rx is $190$ cm away from the SUT to measure the far field. Rx is placed further away from the surface to reduce the coupling between Tx and Rx. In addition, absorbers are also placed on the line of sight between them. All measurements are conducted in the anechoic chamber to reduce background noises and reflections. The output power of the VNA is set to be $10$ dBm.

\begin{figure}[!t]
\centering
\includegraphics[width=3.5in]{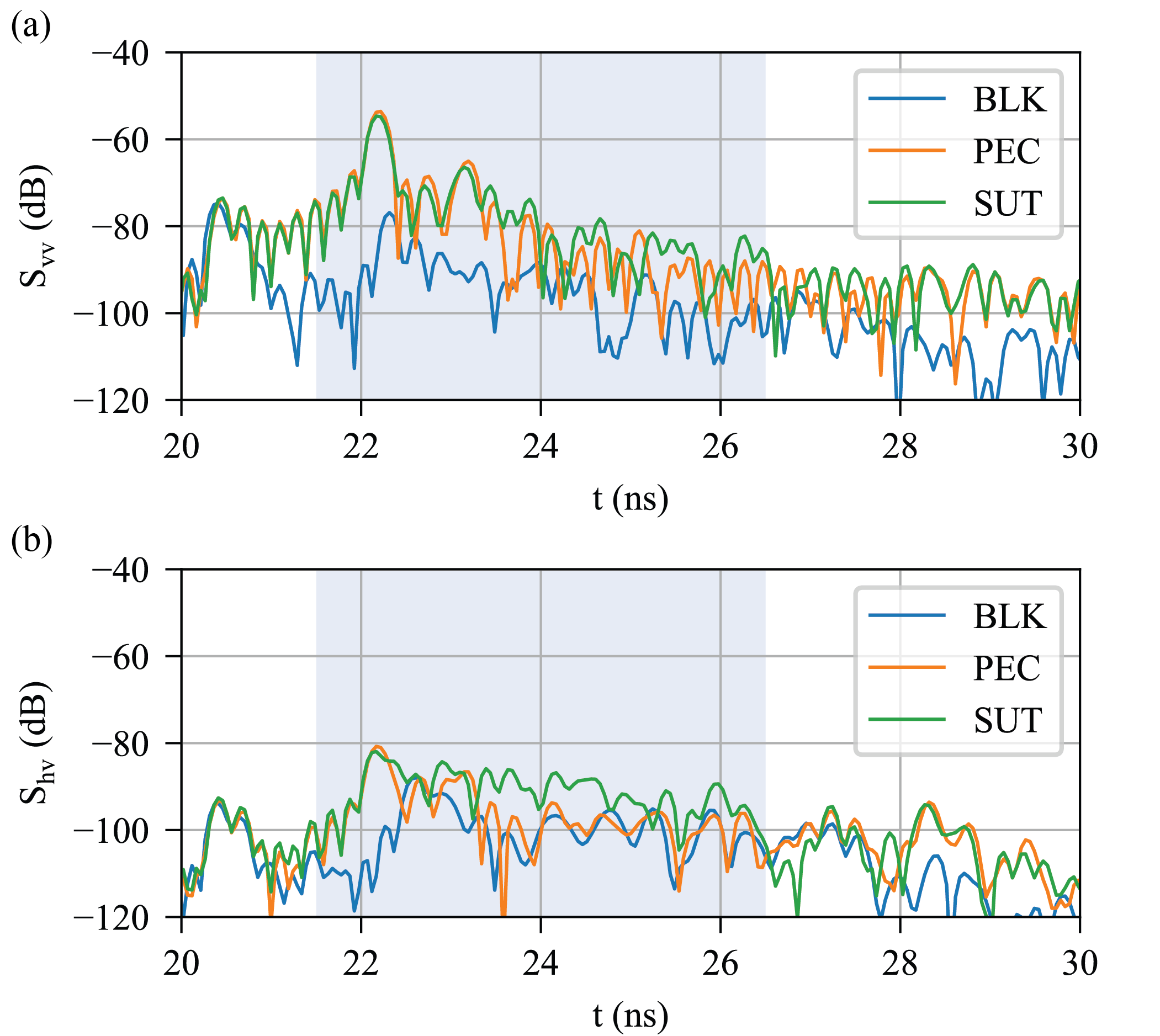}
\caption{Post processing time-gating technique applied to recorded data. (a) and (b): an example of collected $S_{vv}$ and $S_{hv}$ with a rectangular window (shaded area), respectively.}
\label{DP} 
\end{figure}

Three sets of data, with/without SUT and an equal size metal sheet case (aluminum, PEC case), is collected for both the reflection phase and the pattern measurement. The PEC case is used as a baseline study to observe the maximum power that the SUT is able to reflect if lossless. By measuring the blank (BLK) case (without SUT), the data contains all the reflections from the background and the direct talk between Tx and Rx. Thus, subtracting the blank case data from that of the SUT case eliminates environmental signals. In addition, a post-processing technique, such as time-gating, can further filter out the reflected signal from the SUT. A $5$ ns rectangular window is applied to all data after performing Fourier transformation on the collected S parameters, as shown in Fig.\ref{DP}. The gating should start at when the incident wave encounters the SUT, which is the highest peak in the time domain. A $20$ dB signal to noise ratio is observed at the main peak, validating the credibility of the collected data.

\subsection{Measurement Results}
\begin{figure}[!t]
\centering
\includegraphics[width=3.5in]{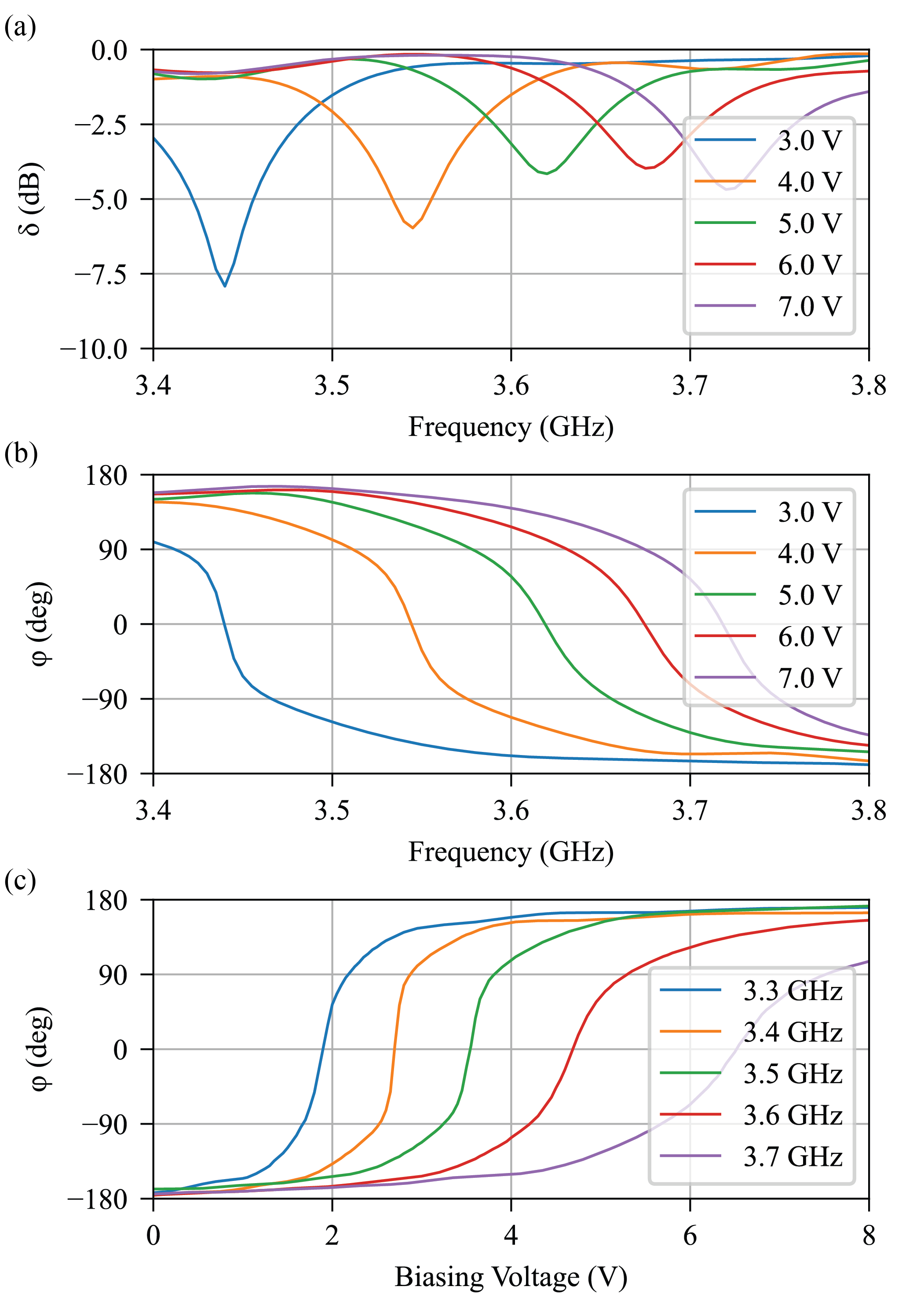}
\caption{Measured S parameter of the metasurface after time-gating and calibration. (a) Loss analysis of the surface. (b) Reflection phase under different biasing voltage. (c) The mapping between biasing voltage and reflection phase.}
\label{Meas.-RP} 
\end{figure}

\begin{figure}[!t]
\centering
\includegraphics[width=3.5in]{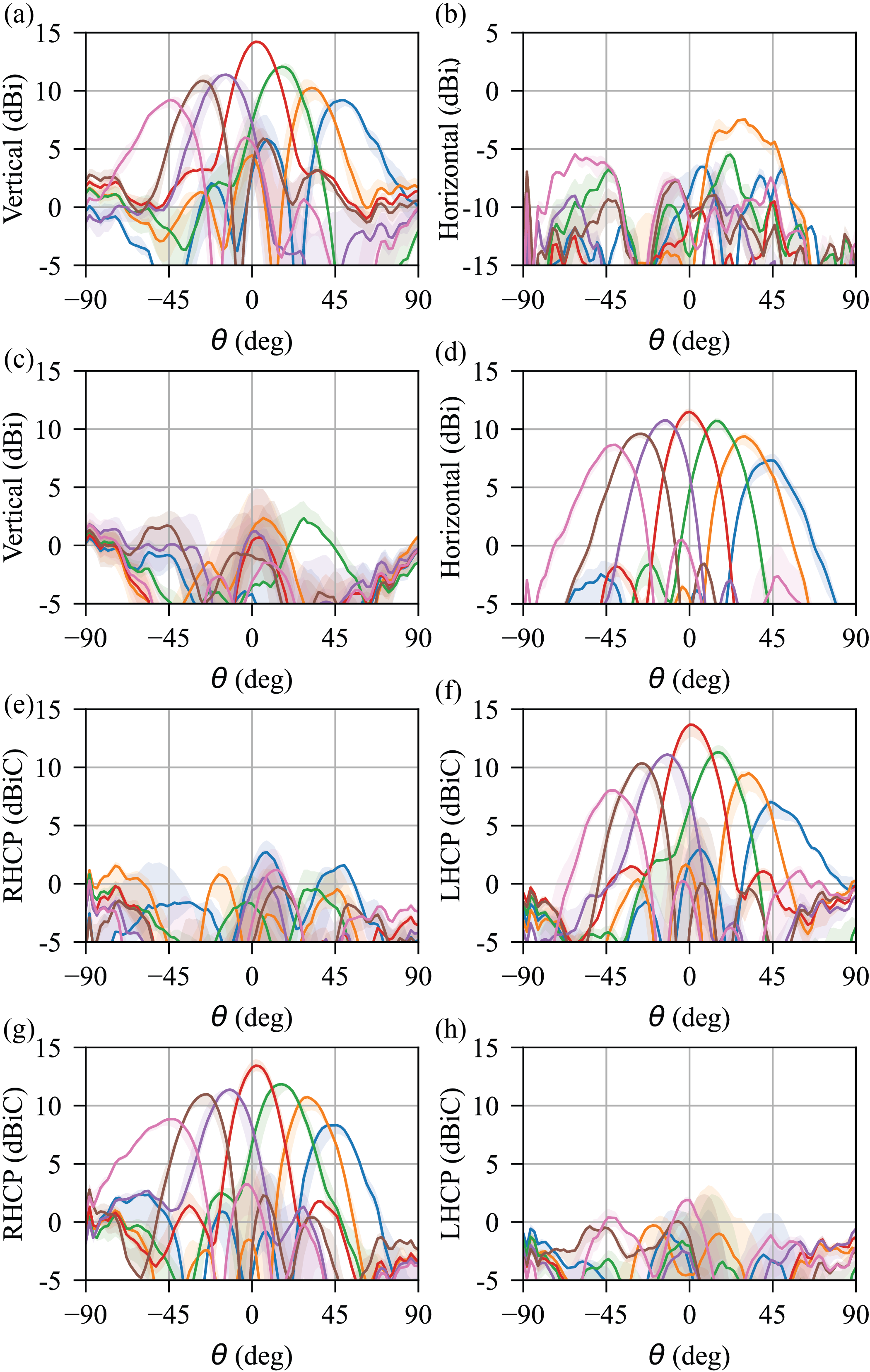}
\caption{Measured pattern for different polarization conversion and steering angle, using the reflection phase look-up table in Fig.\ref{Meas.-RP} (c). (a) and (b): LP to LP conversion. (c) and (d) LP to C-LP conversion. (e) and (f): LP to LHCP conversion. (g) and (h): LP to RHCP conversion. Solid line: $3.6$ GHz, shaded area: $1$ dB bandwidth, $3.58$ GHz to $3.63$ GHz.}
\label{Meas.-Pattern} 
\end{figure}

\begin{figure}[!t]
\centering
\includegraphics[width=3.5in]{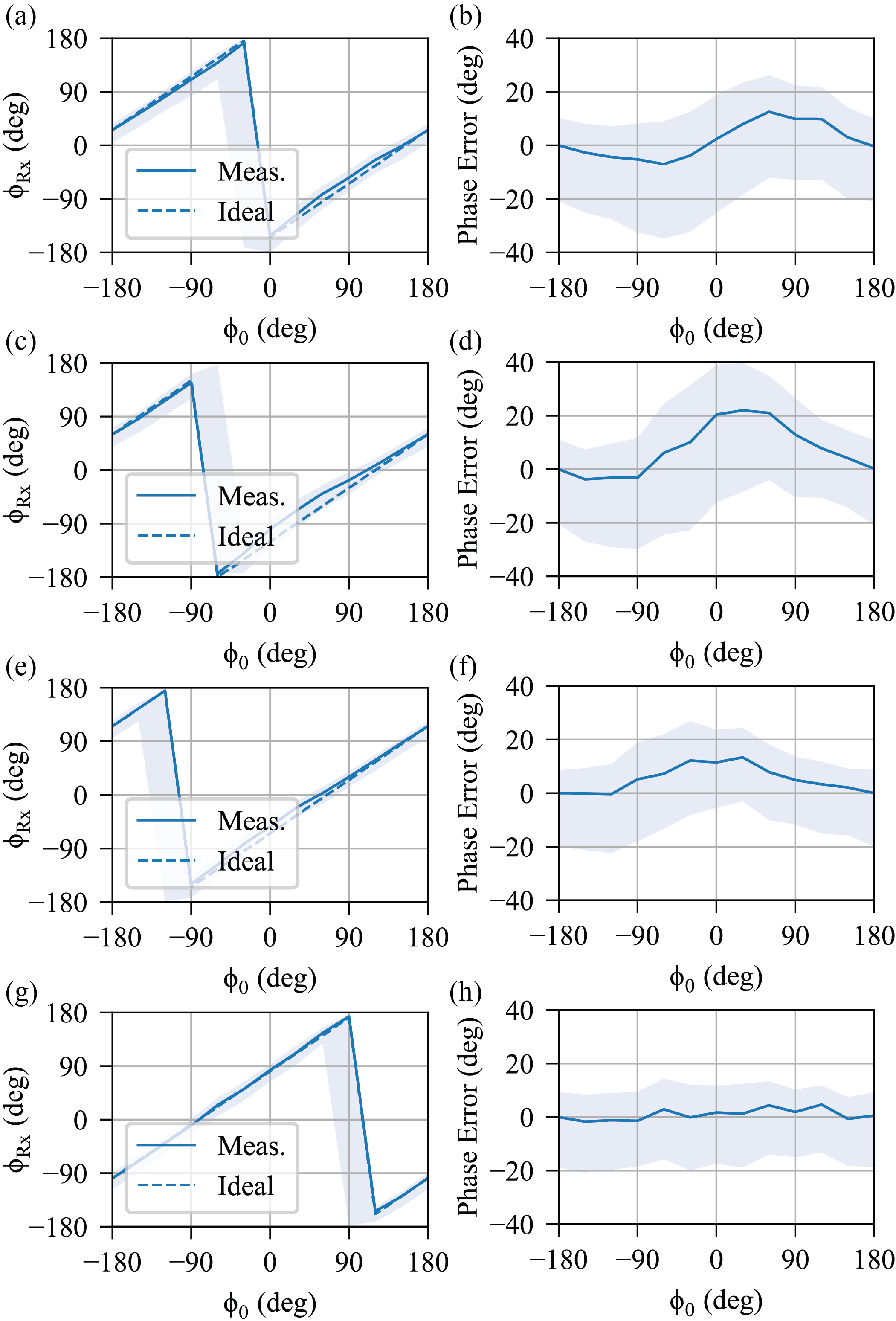}
\caption{The measured phase offset at the main lobe direction by changing the initial phase assigned to the first column of the surface, with a $30^{\circ}$ steering angle. (a) and (b): LP to LP conversion. (c) and (d): LP to C-LP conversion. (e) and (f): LP to LHCP conversion. (g) and (h): LP to RHCP conversion. Solid line: at the main lobe direction. Shaded area: $\pm 5^{\circ}$ of the main lobe direction.}
\label{Meas.-Offset} 
\end{figure}

The first experiment measures the reflection phase of the surface under different biasing voltage. All channels are biased with the same voltage when sweeping from $0$ V to $19$ V at a step of $0.05$ V under normal incidence and the motor at $\theta=0^{\circ}$, as in Fig.\ref{Meas.-Setup} (a). Since polarization conversion is absent in this measurement, $S_{vv}$ for the SUT/BLK/PEC cases measures the loss and reflection phase of the metasurface. The calibrated S parameter of the surface is determined by
\begin{equation}
    S^{SUT,cal}_{vv}(f,u) = \frac{S^{SUT}_{vv}(f,u)-S^{BLK}_{vv}(f,u)}{S^{PEC}_{vv}(f,u)-S^{BLK}_{vv}(f,u)},
    \label{cal}
\end{equation}
where $f$ and $u$ denotes the frequency and biasing voltage for all channels, respectively. Since the equal size PEC case can be considered as lossless, after subtracting the enviromenal noises, dividing the SUT by the PEC case directly indicates the loss ($\delta$) and deembedded reflection phase ($\phi$) of the metasurface,
\begin{equation}
    S^{SUT,cal}_{vv}(f,u) = \delta(f,u)e^{j\phi(f,u)}
\end{equation}
As depicted in Fig.\ref{Meas.-RP} (a), $\delta$ decreases as biasing voltage increases, because the varactors are more lossy at higher capacitance. The maximum loss around $3.6$ GHz is approximately $4$ dB, which is acceptable. A complete phase range from $-180^{\circ}$ to $180^{\circ}$ is observed in Fig.\ref{Meas.-RP} (b). The reflection phase versus biasing voltage curves in (c) demonstrates the tuning bandwidth of the metasurface covers from $3.3$ GHz to $3.7$ GHz, as these frequencies can take any reflection phase within $-180^{\circ}$ to $180^{\circ}$ with a reasonable $\delta$.

The look-up table between biasing voltage and reflection phase in Fig.\ref{Meas.-RP} (c) is used to design the biasing voltages for simultaneous beam steering, polarization conversion and phase offset. A total of $1091$ combinations of $\Delta\phi_i$ (corresponds to different polarization conversion), steering angle and $\phi_0$ (phase of units in the first column), as shown in Table.\ref{case}, is measured for radiation pattern using the setup in Fig.\ref{Meas.-Setup} (a). To save time, the DAQ sweeps all voltage combinations at one azimuth angle before the motor rotates to the next $\theta$.

\begin{table}[!ht]
    \centering
    \caption{Combinations of measured cases.}
    \begin{tabular}{c|ccc}
    \hline
        Parameter & Start & End & Step \\ \hline
        $f$ & $2.5$ GHz & $5.0$ GHz & $0.01$ GHz \\ 
        $\theta$ & $-90^{\circ}$ & $90^{\circ}$ & $2^{\circ}$ \\
        $\Delta\phi_i$ & $-90^{\circ}$ & $180^{\circ}$ & $90^{\circ}$ \\ 
        Steering angle & $-50^{\circ}$ & $50^{\circ}$ & $5^{\circ}$ \\ 
        $\phi_0$ & $-180^{\circ}$ & $180^{\circ}$ & $30^{\circ}$ \\ \hline
    \end{tabular}
    \label{case}
\end{table}

Here, we verify its performance by showing the results at $3.6$ GHz in Fig.\ref{Meas.-Pattern}. Although the data only records $S_{vv}$ and $S_{hv}$, the calibrated S parameter (after using Eqn.\ref{cal} for $S^{SUT}_{vv}$ and $S^{SUT}_{hv}$) between vertical to RHCP/LHCP can be reconstructed by
\begin{equation}
    S^{SUT,cal}_{rv} = \frac{S^{SUT,cal}_{vv}+iS^{SUT,cal}_{hv}}{\sqrt{2}},
\end{equation}
\begin{equation}
    S^{SUT,cal}_{lv} = \frac{S^{SUT,cal}_{vv}-iS^{SUT,cal}_{hv}}{\sqrt{2}}.
\end{equation}

To convert these S parameter into directivity, we perform HFSS simulation for the PEC case under normal plane wave incidence to derive its directivity $D^{PEC}=15.0$ dBi/dBiC. Then, a relationship between $S^{SUT,cal}$ and directivity $D^{SUT}$ is established,
\begin{equation}
    D^{SUT} = S^{SUT,cal} + D^{PEC} = S^{SUT,cal} + 15.0\;\text{dBi/dBiC}.
\end{equation}
The results for four polarization conversion scenarios with different steering angle are shown in Fig.\ref{Meas.-Pattern}. Similar to the simulation results, a cosine envelope is observed across various steering angle, since the effective aperture of the surface is decreasing as the steer angle increases. The cross polarization discrimination (XPD) at $3.6$ GHz in the main lobe direction is present in Table \ref{Table-XPD}. A general isolation above $10$ dB is observed, only $45^{\circ}$ steering LP to LHCP, and $-45^{\circ}$ steering LP to RHCP conversion are below that. Note that this is not because of the large steering angle or conversion to CP, since $-45^{\circ}$ steering LP to LHCP, and $45^{\circ}$ steering LP to RHCP conversion both show above $10$ dB XPD. It happens due to the discrepancies among non-ideal varactors: the accuracy gets worse for certain cases involving more elements working around the resonant frequency. This could be improved by a fine capacitance tuning of each channel, or using machine learning techniques to study the relationship between biasing voltage and reflection phase, which considers the discrepancy of the varactors and manufacturing deviations \cite{Wen23_ML}. However, this is beyond the scope of this paper.

\begin{table}[!ht]
    \centering
    \caption{XPD at the main beam direction (Unit: dB).}
    \begin{tabular}{c|ccccccc}
    \hline
        Converted to & $-45^{\circ}$ & $-30^{\circ}$ & $-15^{\circ}$ & $0^{\circ}$ & $15^{\circ}$ & $30^{\circ}$ & $45^{\circ}$ \\ \hline
        LP & 15.3 & 25.0 & 34.0 & 24.9 & 20.0 & 12.6 & 17.3 \\ 
        C-LP & 18.1 & 12.8 & 26.1 & 11.1 & 13.2 & 16.1 & 25.8 \\ 
        LHCP & 15.0 & 22.3 & 21.4 & 22.6 & 16.9 & 13.0 & 5.5 \\ 
        RHCP & 8.8 & 13.4 & 16.0 & 24.3 & 16.1 & 15.6 & 11.8 \\ \hline
    \end{tabular}
    \label{Table-XPD}
\end{table}

By sweeping the initial phase $\phi_0$ from $-180^{\circ}$ to $180^{\circ}$, the phase of the outgoing beam varies accordingly from $-180^{\circ}$ to $180^{\circ}$. Fig.\ref{Meas.-Offset} illustrates an example of the measured and ideal phase offset with $30^{\circ}$ steering for four polarization conversion scenarios. The phase offset in the measurement is defined by
\begin{equation}
    \phi^{Meas.}_{Rx} = \text{ang} (S_{xv}^{SUT,cal}),
\end{equation}
and the ideal phase is plotted by
\begin{equation}
    \phi^{Ideal}_{Rx} = \phi^{Meas.}_{Rx}(\phi_0 = -180^{\circ},f=3.6\;GHz)+\phi_0.
\end{equation}
The phase error is defined by the difference between $\phi_{Rx}^{Meas.}$ and $\phi_{Rx}^{Ideal}$, and it is shown in the right column of Fig.\ref{Meas.-Offset}. A reasonable phase error ranges within $-20^{\circ}$ to $20^{\circ}$ for all polarization conversions at the main lobe directions, it goes up to $-40^{\circ}$ to $40^{\circ}$ within $\pm 5^{\circ}$ of the main beam direction. This error could also be improved by the techniques mentioned above.

\section{Conclusion} \label{Conclusion}
A four-layer varactor-based multifunctional metasurface for simultaneous beam steering, polarization conversion and phase offset is proposed and experimentally demonstrated for the first time. The incident LP is naturally decomposed by the metallic patches into two orthogonal LPs, whose reflection phase is controlled by the loaded biased varactor. Unlike the N-bit p-i-n diode approach, which is limited by the states of the diodes, this method is capable of continuous phase offset from $180^{\circ}$ to $180^{\circ}$ and is not reported in other works. Although neglecting the differences between varactors and using the overall reflection phase of the whole surface as that of a unit cell, more than $10$ dB XPD is observed for almost all polarization conversion scenarios, LP (X/Y polarization) to LP/C-LP/RHCP/LHCP, from $-45^{\circ}$ to $45^{\circ}$ steering and a $\pm20^{\circ}$ error of phase offset at the main beam direction is achieved. The bandwidth covers from $3.3$ GHz to $3.7$ GHz, while once a combination of biasing voltages is chosen, the $1$ dB bandwidth is $50$ MHz. We believe the proposed metasurface can find applications in wave manipulation, increasing communication efficiency and creating spatial hot/cold spot.

\section*{Acknowledgment}

The authors would like to thank M. Dunna for valuable discussions.


%

\ifCLASSOPTIONcaptionsoff
  \newpage
\fi



%

\end{document}